\documentclass[twocolumn]{aastex701}
\usepackage{xcolor}
\definecolor{connorpurple}{HTML}{5B2C6F}

\usepackage{xcolor}
\usepackage{soul}

\definecolor{connorviolet}{HTML}{4B0082}

\begin{document}

\title[]{The Structural Abundance Crisis of Massive Galaxies in Current Cosmological Simulations}

\author{Hassen M. Yesuf} 
\affiliation{Shanghai Astronomical Observatory, Chinese Academy of Sciences, 80 Nandan Road, Shanghai, 200030,China}
\email{yesufh@shao.ac.cn}

\author{Connor Bottrell}
\affiliation{International Centre for Radio Astronomy Research, University of Western Australia, 35 Stirling Hwy, Crawley, 6009, WA, Australia}
\email{}

\author{Luis C. Ho}
\affiliation{Kavli Institute for Astronomy and Astrophysics, Peking University, Beijing, 100871, China}
\affiliation{Department of Astronomy, School of Physics, Peking University, Beijing, 100871, China}
\email{}

\author{Aaron S. G. Robotham}
\affiliation{International Centre for Radio Astronomy Research, University of Western Australia, 35 Stirling Hwy, Crawley, 6009, WA, Australia}
\email{}

\author{Sabine Bellstedt}
\affiliation{International Centre for Radio Astronomy Research, University of Western Australia, 35 Stirling Hwy, Crawley, 6009, WA, Australia}
\email{}

\author{Robin H. W. Cook}
\affiliation{International Centre for Radio Astronomy Research, University of Western Australia, 35 Stirling Hwy, Crawley, 6009, WA, Australia}
\email{}

\author{Lei Hao}
\affiliation{Shanghai Astronomical Observatory, Chinese Academy of Sciences, 80 Nandan Road, Shanghai, 200030,China}
\email{}

\author{Fengshan Liu}
\affiliation{National Astronomical Observatories, Chinese Academy of Sciences, 20A Datun Road, Chaoyang District, Beijing, 100101, China}
\email{}

\begin{abstract}

The internal structure of galaxies encodes the complex baryon cycle driven by gas accretion, star formation, and feedback, together with secular evolution and environmentally driven processes such as mergers and tidal interactions. We present a population-level census of massive nearby galaxies ($\log(M_\star/M_\odot) > 10$) by comparing Hyper Suprime-Cam Subaru Strategic Program observations with matched mock images from IllustrisTNG, EAGLE, and SIMBA. Using a consistent, like-for-like imaging pipeline, we construct structural abundance functions (SAFs) for key morphological parameters, revealing a structural abundance crisis. Across S\'{e}rsic index, concentration, size, and ellipticity, all simulations exhibit large, systematic discrepancies ($>5\sigma$; RMSE $\sim 0.2$--$1.8$\,dex), typically corresponding to abundance differences of factors of several. While individual simulations display diverse failures---underproducing or overproducing compact spheroids, extended or round galaxies ---all underproduce highly flattened disks. Although TNG shows the closest agreement and SIMBA the largest offsets, this shared failure indicates that current models---despite matching global demographics such as the stellar mass function---do not uniquely constrain internal galaxy structure. Our results demonstrate that agreement in integrated observables can mask fundamental shortcomings in the modelling of mass and angular momentum redistribution. We therefore establish SAFs as a stringent, multidimensional, and observationally accessible benchmark for testing and calibrating next-generation galaxy formation models in the era of upcoming deep, wide-field surveys.
\end{abstract}



\section{Introduction}

Reproducing the structural diversity of galaxies remains one of the most formidable challenges for galaxy formation models within the $\Lambda$CDM framework \citep{PeeblesNusser10}. While state-of-the-art cosmological hydrodynamical simulations---most notably EAGLE \citep{Schaye+15}, IllustrisTNG\citep{2018MNRAS.475..648P,2018MNRAS.475..624N, Nelson2019MNRAS.490.3234N,Pillepich+19} and SIMBA \citep{Dave+19}---broadly reproduce global demographics such as stellar mass functions and cosmic star formation histories \citep{2018MNRAS.475..648P, Dave+19, Furlong+15, Katsianis+17}, these integrated benchmarks mask a systemic structural abundance discrepancies. Agreement in global quantities can conceal divergent underlying physics: similar stellar mass functions may arise from substantially different implementations of star formation and feedback, whose discrepancies become evident only when we examine subpopulation statistics \citep{YesufBottrell26} and internal stellar distributions.

The internal structure of galaxies provides a sensitive and cumulative record of their full formation and assembly history. It encodes the interplay between gas accretion, mergers, secular evolution, star formation, and feedback-driven outflows, all of which redistribute mass and angular momentum on different spatial scales and timescales. These processes collectively shape galaxy structure along a continuous distribution of dynamical states and morphologies, including rotation-dominated disk galaxies with a wide range of bulge contributions, as well as dispersion-dominated systems that span both compact and extended spheroidal configurations. 

At early cosmic times ($z \gtrsim 2$), high gas accretion rates and frequent dissipative events drive intense star formation and strong feedback activity, making both stellar and AGN feedback central to shaping galaxy structure. At later times ($z \lesssim 2$), as accretion rates decline and galaxies become increasingly gas-poor, the relative importance of these processes shifts, with dry and mixed mergers playing a dominant role in the structural evolution of massive galaxies \citep{Huang+13,Huang+16,Davari+17} while feedback processes primarily regulate the gas supply and suppress further in-situ star formation. 

In massive galaxies, active galactic nucleus (AGN) feedback plays a key role in shaping the observed morphological mix by quenching in-situ star formation and suppressing gas accretion, thereby limiting disk regrowth and increasing the importance of ex-situ stellar assembly through gas-poor mergers. This evolutionary pathway promotes the formation and maintenance of bulge-dominated systems and contributes to the emergence of extended quenched galaxies at the massive end \citep{Huang+13,Dubois+16,Huang+16,Davari+17,Weinberger+17,ChoiE+18}. 

In lower-mass galaxies, stellar feedback regulates star formation and the baryonic angular momentum distribution---preferentially removing low-angular-momentum gas and driving turbulence in the interstellar medium---thereby supporting extended disk formation \citep{Hopkins+14,ElBadry+18}. In addition, secular evolution, mediated by bars and spiral arms, further redistributes angular momentum and shapes galaxy structure \citep{KormendyKennicutt04}. In non-isolated systems, galaxy structure is influenced by environment-dependent external perturbations.

The diversity of galaxy structure thus reflects a delicate balance between these processes and their evolution over time. Sustained accretion of high-angular-momentum gas promotes disk growth \citep{Grand+17, Tacchella+19}, whereas early, dissipative star formation or inefficient feedback regulation leads to excessively compact, centrally concentrated systems. Because these processes operate on scales below the resolution of cosmological simulations, models rely on calibrated subgrid prescriptions. Even minor variations in these implementations---under otherwise identical initial conditions---can produce markedly different structural outcomes \citep{Scannapieco+12,CrainvdVoort23}. Consequently, the full distribution of galaxy structures provides a stringent test of whether simulations capture the physics of baryonic assembly, redistribution, and galaxy-scale dynamical evolution.

In this work, we move beyond average scaling relations to present a population-level census of galaxy structure. We construct a robust observational benchmark of structural abundance functions (SAFs), defined as the differential number densities of galaxies as a function of structural parameters, analogous to the stellar mass function but extended to multiple indices of galaxy structure. We further present conditional SAFs subdivided by stellar mass and star-formation state (star-forming and quiescent). To enable a direct, like-for-like comparison, we combine deep imaging from the Hyper Suprime-Cam Subaru Strategic Program \citep[HSC;][]{Aihara+18} with highly complete spectroscopy from the Galaxy And Mass Assembly survey \citep[GAMA;][]{Driver+22} and process mock observations from IllustrisTNG, EAGLE, and SIMBA through radiative transfer and imaging pipelines, reproducing the noise and resolution characteristics of the HSC data \citep{Bottrell+24}.

While previous studies have demonstrated that state-of-the-art simulations broadly reproduce several average structural trends, they have also identified important tensions in galaxy sizes, morphologies, and disk thickness \citep{Furlong+17, Clauwens+18, Genel+18, Tacchella+19, Rodriguez-Gomez+19, Dave+19, Bignone+20, deGraaff+22, XuD+24, Eisert+26}. However, these comparisons have primarily focused on scaling relations or the distributions of individual structural properties, leaving the population-level abundance of different galaxy structures largely unexplored. Here, by constructing SAFs across multiple structural parameters and stellar populations, we provide a complementary census of galaxy structure analogous to the stellar mass function. Applying this framework consistently to three flagship simulations reveals systematic differences in the abundance of compact spheroids, extended disks, and other structural populations. Including SIMBA---whose structural properties have been less extensively explored---extends previous comparisons beyond the TNG and EAGLE models that have dominated earlier studies and enables us to assess whether structural discrepancies arise from shared limitations across different galaxy formation simulations. We present SAFs as a new benchmark for evaluating whether galaxy formation models reproduce not only average galaxy properties but also the observed diversity and relative abundance of galaxy structures.

\section{Methods}

\subsection{Simulations}

We compare our observational measurements with three state-of-the-art cosmological hydrodynamical simulations: IllustrisTNG100 \citep{2018MNRAS.480.5113M,2018MNRAS.477.1206N,2018MNRAS.475..624N,2018MNRAS.475..648P,2018MNRAS.475..676S}, IllustrisTNG50 \citep{Nelson2019MNRAS.490.3234N,Pillepich+19}, EAGLE \citep{2015MNRAS.450.1937C,Schaye+15}, and SIMBA \citep{Dave+19}. Each simulation employs a different numerical approach---moving-mesh (TNG), smoothed particle hydrodynamics (EAGLE), and meshless finite mass (SIMBA)---and incorporates distinct subgrid models for star formation, stellar feedback, black hole growth, and AGN feedback \citep{Weinberger+17}. These differences provide complementary predictions for galaxy structure, and stellar mass distribution, and all simulations are analyzed using methods consistent with the observational dataset.

\subsubsection{IllustrisTNG}

IllustrisTNG is a suite of cosmological magneto-hydrodynamical simulations run with the moving-mesh code AREPO \citep{2010MNRAS.401..791S}. For this work, we use TNG100 ($L = 75\,h^{-1}$\,Mpc) to capture representative large-scale volumes, and TNG50 ($L = 35\,h^{-1}$\,Mpc) to probe the effects of higher spatial and mass resolution. In TNG100, the dark matter and baryon mass resolutions are $7.5\times10^6\,M_\odot$ and $1.4\times10^6\,M_\odot$, respectively, while TNG50 reaches $4.5\times10^5\,M_\odot$ and $8.5\times10^4\,M_\odot$. Star formation follows a Kennicutt–Schmidt-like relation in dense gas ($n_\mathrm{H} > 0.1\,\mathrm{cm}^{-3}$), and stellar feedback is implemented as kinetic winds whose velocity scales with local dark matter velocity dispersion and whose mass-loading is metallicity-dependent. Wind velocities also evolve with redshift to maintain the specified energy per stellar mass formed.

Black holes are seeded in halos above $5\times10^{10}\,M_\odot$ and grow via mergers and Bondi–Hoyle accretion capped at the Eddington limit. AGN feedback transitions between thermal and kinetic modes depending on the accretion rate: high-accretion thermal feedback injects energy isotropically, while low-accretion kinetic feedback launches directed momentum kicks that are isotropic on average over time. Subgrid parameters were calibrated to reproduce the $z=0$ galaxy stellar mass function, global star formation rate density, and stellar–halo mass relation, while also broadly reproducing black hole scaling relations, halo gas fractions, and galaxy sizes. 

\subsubsection{EAGLE}

EAGLE Ref-L100N1504 simulates dark matter and baryonic particles in a $68\,h^{-1}\,\mathrm{Mpc}$ box using a pressure-entropy formulation of SPH. Its dark and baryonic particle resolutions are $9.7\times10^6\,\mathrm{M}_\odot$ and $1.8\times10^6\,\mathrm{M}_\odot$, respectively. Star formation is stochastic, governed by a metallicity-dependent density threshold, and scales with local gas pressure in a Kennicutt–Schmidt-like manner. Stellar feedback is implemented through stochastic thermal heating, in which young stellar populations heat neighboring gas particles by a fixed temperature increment, with efficiency depending on local gas properties.

Black holes are seeded in halos above $10^{10}\,M_\odot$ and grow via mergers and modified Bondi accretion that accounts for angular momentum suppression, capped at the Eddington limit. AGN feedback is delivered thermally and stochastically, with energy accumulated and released once sufficient energy is available to heat neighboring gas. The subgrid model was primarily tuned to reproduce the $z=0$ stellar mass function and galaxy size–mass relation, and the AGN efficiency was adjusted to match observed black hole–stellar mass scaling.

\subsubsection{SIMBA}

SIMBA is run with the meshless finite mass code GIZMO in a $100\,h^{-1}$\,Mpc box (for m100n1024 run), with dark matter and baryonic particle masses of $9.6\times10^7\,M_\odot$ and $1.8\times10^7\,M_\odot$. Star formation follows a molecular Kennicutt–Schmidt relation, while stellar feedback is implemented via two-phase winds carrying metals and dust. Wind velocities scale with galaxy circular velocity, and approximately 30\% of wind particles are launched hot at $T=10^{5.5}$\,K.

Black holes are seeded in galaxies above a mass threshold, accreting via torque-limited (cold gas) or Bondi (hot gas) modes, both capped at the Eddington rate. AGN feedback operates through kinetic winds and jets, with the dominant mode determined by Eddington ratio and black hole mass, and includes secondary X-ray heating in low-gas systems. Subgrid parameters are calibrated primarily to reproduce the $z=0$ stellar mass function, with black hole accretion efficiencies tuned to match the observed black hole–stellar mass relation.

\subsection{Generating Mock HSC Images from Simulations}

We select galaxies with $\log(M_\star/M_\odot) > 10$ at $z \approx 0.1$ from each simulation for radiative transfer post-processing and injection into HSC background images (snapshot 91 for TNG, 27 for EAGLE, 145 for SIMBA) \citep{Bottrell+24, Eisert+24,Eisert+26}. Stellar emission is modeled with dust attenuation using the SKIRT Monte Carlo radiative transfer code \citep{CampsBaes20,Bottrell+24}, which simulates the propagation of starlight through a dusty interstellar medium, including absorption, scattering, and thermal re-emission, to produce physically realistic, noise-free images.

These idealized images are subsequently processed with a customized version of RealSim \citep{Bottrell+19}, where they are convolved with reconstructed HSC point-spread functions, rebinned to the CCD pixel scale (0.168 arcsec), and inserted into real final-depth HSC Wide cutouts. Insertion locations are drawn from an HSC photometric redshift catalogue with well-characterized sources \citep{Tanaka+18}, ensuring that the depth, seeing, and crowding properties of the synthetic images match those of the observations. This procedure yields mock images with realistic surface brightness, morphology, and photometric characteristics directly comparable to the data.

Mock observations are generated for 11,963 EAGLE, 21,564 TNG100 and 3,073 TNG50 galaxies using four tetrahedrally arranged viewing angles. The tetrahedron has fixed sightlines with respect to the box, with one sightline along the z-axis of the simulation volume. For 10,976 SIMBA galaxies the single z-axis sightline is used due to larger data size. All simulations are processed through the same pipeline, including dust radiative transfer, stellar emission modeling, and insertion into HSC fields, providing a consistent, like-for-like comparison. Full details of the image-generation procedure are provided in \citet{Bottrell+24}.

At the median redshift of the HSC sample ($z \sim 0.1$), the effective HSC spatial resolution set by the typical seeing corresponds to $\sim1$\,kpc. By comparison, the native stellar gravitational softening lengths of the simulations are $\sim0.3$\,kpc for TNG50 and $\sim0.7$\,kpc for TNG100, EAGLE, and SIMBA. The simulated galaxies are forward-modelled with realistic HSC seeing, pixelization, and sky backgrounds, thereby controlling the observational component of the comparison while leaving possible residual effects of native spatial and mass resolution.

Although the common forward modelling pipeline places all four simulations on the same HSC PSF, pixel-scale, and background-noise distributions, it homogenizes observational image quality rather than native mass and force resolution. At fixed \(M_\star\), SIMBA's nominal baryonic resolution elements are \(\sim210\) times more massive than those of TNG50; mass resolution sets how many elements sample compact and extended components, while force resolution limits the smallest self-gravitating structures that can form. Sparse sampling can therefore affect both the evolved stellar distribution and its rendering into light, and subsequent PSF convolution and noise can further dilute compact nuclei and faint outskirts. Controlled comparisons of ideal and survey-degraded images show that image quality and particle-to-light smoothing can alter recovered structure, particularly bulge-sensitive quantities \citep{2017MNRAS.467.1033B,2019MNRAS.486..390B,2021MNRAS.507..886T}; related merger diagnostics are likewise sensitive to depth and resolution \citep{2024MNRAS.528.5558W,2024MNRAS.534.2533B}.

\citet{2017MNRAS.467.2879B} also found that photometric and kinematic bulge fractions diverge increasingly toward lower stellar mass. This is a bidirectional limitation: sparse particle sampling can compromise dynamical component separation, whereas poor seeing can conceal a compact photometric component. We therefore expect SIMBA to be most susceptible to resolution-related systematics, generally at lower mass and for central-concentration diagnostics such as \(n\), although neither the magnitude nor sign of the bias need be common to all statistics.

\subsection{Observational Data: Subaru HSC Wide Survey and GAMA}

We use imaging from the Hyper Suprime-Cam (HSC) Subaru Strategic Program Wide layer \citep[HSC-SSP Wide;][]{Aihara+18, Aihara+22}, which provides deep, multi-band optical imaging in $grizy$ filters. At final depth, the Wide layer reaches $5\sigma$ point-source depths of $i \sim 26$ with typical seeing of $0.6^{\prime\prime}$ and a pixel scale of 0.168 arcsec, enabling robust structural measurements. We focus on fields overlapping with the Galaxy And Mass Assembly (GAMA) equatorial regions, covering $\sim 126$\,deg${^2}$. Galaxy structural parameters are measured in all bands, independently. Our analysis focuses on the $i$-band measurements which which provide the highest signal-to-noise and best seeing.

Galaxy stellar masses, star formation rates (SFRs), and redshifts are drawn from the GAMA Data Release 4 \citep[DR4;][]{Driver+22}, a magnitude-limited spectroscopic survey covering the equatorial HSC fields. Stellar masses and SFRs are derived via spectral energy distribution fitting to multi-band photometry, in some cases spanning from the far-ultraviolet to the far-infrared \citep{Bellstedt+20}. The combination of HSC imaging and GAMA spectroscopy yields a sample with high-quality structural measurements and reliable stellar masses, enabling robust comparisons with simulated galaxy populations.

We select galaxies from the HSC--GAMA overlap in the G09, G12, and G15 fields by requiring reliable $i$-band structural measurements, redshifts in the range $0.015 < z \leq 0.12$, and stellar masses $\log(M_\star/M_\odot) \geq 10$. To ensure uniform coverage, we further restrict the sample to the common survey footprint using cuts in right ascension and declination ($\mathrm{Dec} > -1.5$, with field-dependent upper limits; $\mathrm{Dec} < 1.0$ for G15). The resulting sample comprises 6{,}348 galaxies and is $\sim$90\% complete.

A small fraction of galaxies (0.2--0.3\%) are excluded owing to unsuccessful structural fits, with a comparable rejection rate in both observations and simulations. These failures arise almost exclusively from cases in which the source-detection algorithm fails or is flagged. Specifically, we generate large image cutouts—scaled from the original GAMA \textsc{ProFound} apertures—and re-run source detection on the HSC imaging. In rare instances, the algorithm identifies an extended region of the field as a single object, typically due to contamination from bright stars blended with the source. Such cases are removed from the analysis.

\subsection{Structural Measurements}

We quantify galaxy structure using both parametric and non-parametric diagnostics. Source detection and photometry are performed with \textsc{ProFound} \citep{Robotham+18}, and structural parameters are extracted using \textsc{ProFuse} \citep{Robotham+22}. Source detection and deblending are carried out on a co-added stack of all HSC $grizy$ bands along with the corresponding variance maps. Photometry is then measured independently in each band using the stack segmentation image, ensuring that the highest possible signal-to-noise is used for source detection and deblending, even for galaxies that are faint in individual bands. 

Galaxy light profiles are modeled with \textsc{ProFit} \citep{Robotham+18}, fitting a seeing-convolved, single-component S\'{e}rsic profile independently in each band. The HSC Software Pipeline \citep{2018PASJ...70S...5B} models localized point-spread functions (PSF) from isolated stars and the spatial variation between these points, allowing approximate PSFs to be reconstructed for any local in each band. From these fits, we derive the half-light radius ($R_{50}$), projected axis ratio ($b/a$), and S\'{e}rsic index ($n$). The S\'{e}rsic index characterizes the overall shape of the surface brightness profile, capturing sensitivity to both inner components and extended outer wings, whereas $b/a$ traces the projected galaxy shape.

We also compute non-parametric structural indicators, including the concentration index $C_{82} = 5 \log_{10}(R_{80}/R_{20})$, which traces the relative distribution of light between inner and outer regions, and the asymmetry ($A$), which quantifies deviations from rotational symmetry and is sensitive to mergers, disturbances, lopsidedness, and clumpy spiral structure \citep{Conselice03, Yesuf+21}. The Petrosian radius, $R_{\rm Petro}$, is defined as the radius at which $\eta(R_{\rm Petro}) = 0.2$, where $\eta(R) = I(R) / \langle I (<R) \rangle$, $I(R)$ is the surface brightness profile at radius $R$, and $\langle I(<R) \rangle$ is the average surface brightness within $R$.

All parameters are measured from $i$-band imaging using an identical analysis pipeline for both observations and simulations, applied to HSC data and to mock HSC images constructed from the simulations. This forward-modelling approach ensures a consistent, like-for-like comparison by subjecting both datasets to the same observational effects and measurement systematics. Consequently, any differences between observed and simulated galaxy populations reflect intrinsic physical discrepancies rather than methodological biases.

The S\'{e}rsic index and concentration both probe galaxy light distributions but differ in their sensitivities. Concentration provides an integrated measure based on enclosed flux, whereas the S\'{e}rsic index captures the detailed surface brightness profile shape and is more sensitive to inner components, such as bulges, and extended outer features. We find that concentration distributions are more consistently reproduced across simulations, while S\'{e}rsic index shows larger systematic differences, reflecting its sensitivity to small-scale structural variations. Together, these diagnostics provide complementary constraints on galaxy structure.

For the simulations, stellar masses and SFRs represent the intrinsic properties of the model galaxies. Specifically, we use the stellar mass and the instantaneous SFR within twice the half-mass radius provided by the simulations. Observational stellar masses in GAMA are derived from SED fitting and can differ systematically from the intrinsic simulation masses. Although mock SED fitting can provide more direct observational analogues, it is not essential for our analysis because we adopt broad stellar-mass bins and coarse star-formation categories, which reduce the impact of moderate mass offsets. Moreover, the global stellar mass functions of the simulations broadly agree with the observations when compared using coarse mass bins.

\subsubsection{Star-forming and Quiescent Categories}

Star-forming galaxies exhibit a tight correlation between stellar mass ($M_\star$) and star formation rate (SFR), commonly referred to as the star formation main sequence (SFMS). Following previous work, we adopt a linear relation $\log \mathrm{SFR} = \alpha \log M_\star + \beta$ with slope $\alpha = 0.8$ and intercept $\beta = -8.1$ \citep{YesufBottrell26}.

We classify galaxies as star-forming if their offset from the SFMS satisfies $\Delta \log \mathrm{SFR} > -1$\,dex, where $\Delta \log \mathrm{SFR}$ is defined relative to the mean SFMS relation above. This definition includes green-valley galaxies.

High-S\'{e}rsic-index ($n > 4$) star-forming galaxies in the HSC sample are preferentially located below the SFMS, with a median (16th, 84th percentile) offset of $\Delta \log \mathrm{SFR} = -0.4\,$(--0.8, 0.02)\,dex. Visual inspection indicates that, in many of these systems, star formation is primarily confined to the outer regions, while the central regions are largely quiescent. The inspection further confirms that these objects are genuine galaxies rather than unobscured AGN contaminants with bright central point sources; such contaminants constitute a negligible fraction of the sample ($<2\%$).

\subsubsection{Correcting for Observational Bias and Incompleteness: $1/V_\mathrm{max}$ Method}

We correct for Malmquist bias using the $1/V_\mathrm{max}$ method \citep{Pozzetti+10, Baldry+12, YesufBottrell26}, estimating number densities as a function of galaxy structural parameters. For each galaxy, the maximum observable redshift, $z_\mathrm{max}$, is derived from the stellar mass completeness limit using 95\% quantile regression, performed separately for star-forming and quiescent populations to account for their systematically different $M/L$ ratios. These mass completeness limits are translated from the survey flux limit ($r < 19.7$ for GAMA) using $k$-corrections from color–redshift relations \citep{Chilingarian+10} and dust attenuation corrections \citep{Calzetti+00}.  

The maximum comoving volume accessible to each galaxy, $V_\mathrm{max}$, is then computed using $z_\mathrm{max}$, and all number densities are weighted by $1/V_\mathrm{max}$ to correct for the over-representation of intrinsically luminous or massive objects. Given the depth of HSC imaging, structural parameters are available for essentially all galaxies in the HSC–GAMA overlap region, and the same $1/V_\mathrm{max}$ weights are applied to each structural measurement.

The observational uncertainties of the logarithmic number densities are obtained by Poisson propagation of the weighted $1/V_{\max}$ estimator, $\sigma_{\log\phi, \mathrm{obs}}=(\ln10)^{-1}[\sum_j w_j^2]^{1/2}/\sum_jw_j$, where $w_j=1/V_{\max,j}$. For the simulations, the uncertainties are estimated from Poisson counting statistics, $\sigma_{\log\phi,\, \mathrm{sim}}=(\ln10)^{-1}/\sqrt{N}$, where $N$ is the number of galaxies in the corresponding structural-parameter and stellar-mass bin under the SAF selection. To quantify the statistical significance of the differences between the observed and simulated structural-parameter distributions, we compare the number densities in corresponding bins using a $\chi^2$ statistic and the root-mean-square error (RMSE). For each bin $i$, we combine the observational and simulation uncertainties in quadrature, $\sigma_i^2=\sigma_{\log\phi, \mathrm{obs},i}^2+\sigma_{\log\phi, \mathrm{sim},i}^2$, assuming that the two statistical errors are independent, and calculate $\chi^2=\sum_i(\log \phi_{\mathrm{obs},i}-\log \phi_{\mathrm{sim},i})^2/\sigma_i^2$. Because no parameters are fitted when comparing the two distributions, the number of degrees of freedom is equal to the number of independent bins. We convert the corresponding $\chi^2$ survival probability to an equivalent one-sided Gaussian significance, which we quote as the number of $\sigma$. These significances quantify statistical discrepancies under the adopted independent-bin error model and do not include additional systematic uncertainties, such as cosmic variance.

The SAFs represent differential number densities of galaxies within fixed stellar mass bins. Logarithmically scaled structural parameters, including concentration, are measured per dex, whereas linear parameters, such as the S\'{e}rsic index, are measured per unit structural parameter. For simplicity, the figure axes are labeled with the number-density units of the SAFs, $\mathrm{Mpc}^{-3}$.

\subsubsection{Cosmic Variance Assessment} 

Our analysis of the conditional structural abundance functions uses a $126~\mathrm{deg}^2$ GAMA+HSC survey area, combining three independent sub-regions (30, 42, and 54~$\mathrm{deg}^2$). We evaluated simulation--observation agreement using the MAE and RMSE independently in each sub-region for both the S\'{e}rsic index and Petrosian radius abundance functions. The field-to-field MAE variation is typically $\sim0.1$--$0.15$\,dex (up to $\sim0.2$\,dex for rare populations), and the integrated GAMA number-density normalization varies by a factor of $\sim1.6$ across fields (always $<2\times$, with the largest variation occurring for rare massive star-forming galaxies). The \citet{DriverRobotham2010} cosmic variance calculator\footnote{\url{https://cosmocalc.icrar.org}} predicts $\pm15\%$ for the full area, confirming that our sub-field estimates provide a conservative upper limit on the sensitivity of the SAFs to cosmic variance. Because the simulations considered here are single-box realizations, we cannot quantify their cosmic variance directly from independent simulation volumes; instead, we estimated it using the empirical relation of \citet{DriverRobotham2010}. Inflating the Poisson errors by factors of $1.15$--$1.35$ to account for cosmic variance (with $1.35$ corresponding to the TNG50 volume) does not change the main conclusions. However, for certain parameters---such as ellipticity subsets, where TNG50 shows lower significance, or massive star-forming galaxies in the Petrosian-radius SAFs for EAGLE and SIMBA---a minority of subsets yield discrepancies of $\sim1$--$4\,\sigma$. Cosmic variance is thus unlikely to be the dominant source of the simulation--observation discrepancies, although it may remain relevant for rare populations. Multiple independent realizations of large cosmological volumes and mock light-cone analyses will enable a more robust quantitative assessment. Upcoming wide-area surveys will cover areas roughly two orders of magnitude larger than GAMA, substantially suppressing cosmic-variance fluctuations and greatly reducing the associated observational uncertainty.

\section{Results}

A population-level comparison reveals that the TNG, EAGLE, and SIMBA simulations deviate systematically from observations, and from one another, across all structural metrics. No simulation simultaneously reproduces the distributions of S\'{e}rsic index ($n$), size ($R_\mathrm{Petro}$, $R_{50}$), ellipticity ($\epsilon$), asymmetry ($A$), and concentration ($C_{82}$) over the full range of stellar masses and star-formation classes. These multi-parameter discrepancies, significant in the vast majority of subsamples ($\gtrsim 5\sigma$), affect both the normalization and shapes of the structural distributions, indicating a broad failure to capture the diversity of galaxy structure.

Matching the global stellar mass function (SMF) is a key calibration requirement for galaxy formation models, particularly for constraining subgrid physics, but it is not sufficient to reproduce realistic galaxy structures, which reflect the internal mass distribution. Moreover, the simulations fail to recover the observed SMF when subdivided into star-forming and quiescent populations \citep{YesufBottrell26}. Although this analysis is conducted in narrow stellar mass bins, discrepancies in SMF normalization and structural abundance can remain coupled, in some cases potentially reflecting a common origin. Crucially, the shapes of the structural distributions in the simulations are generally themselves inconsistent with observations. We now examine how these discrepancies manifest across individual structural metrics.

The S\'{e}rsic index distributions (Fig.~1) reveal a strong tension in the balance between spheroid- and disk-dominated galaxies. Low S\'{e}rsic indices ($n \sim 1$) trace exponential disks, whereas higher values reflect enhanced central concentrations and extended wings associated with prominent bulges and stellar halos. All simulations exhibit significant ($\gtrsim 5\sigma$) systematic offsets, with number-density discrepancies reaching up to one to two orders of magnitude at the high-$n$ end. EAGLE and SIMBA strongly underproduce spheroidal galaxies across all populations, whereas TNG50 shows an excess high-$n$ star-forming galaxies (SFGs) and TNG100 favors disk-like morphologies in quiescent systems. Quantitatively, the TNG suite provides the closest overall agreement (RMSE $\sim 0.1$--$0.7$\,dex), outperforming EAGLE ($\sim 0.4$--$1.3$\,dex) and SIMBA ($\sim 0.8$--$1.8$\,dex). 

Systematic offsets persist in the size distributions (Figs.~2 and~3). Although the three simulations broadly reproduce the mean size--mass relations established in earlier work \citep{Furlong+17, Genel+18, Rodriguez-Gomez+19, Dave+19, deGraaff+22}, none captures the full observed diversity of galaxy sizes. The discrepancies are highly significant, corresponding to tensions of $>5\sigma$ in nearly all mass bins. For Petrosian radii ($R_\mathrm{Petro}$), RMSE span 0.2–0.9\,dex, with TNG50 and SIMBA showing the largest deviations in some subpopulations ($\sim 0.7-0.9$\,dex). All simulations underproduce compact star‑forming galaxies and overproduce extended ones, with the latter trend being most severe in TNG and EAGLE for $R_\mathrm{Petro}$. Conversely, SIMBA underproduces large star‑forming systems when sizes are measured with the half‑light radius ($R_\mathrm{50}$), whereas EAGLE reproduces $R_\mathrm{50}$ of intermediate‑mass star‑forming galaxies relatively well. For quiescent galaxies, TNG and SIMBA overproduce large systems ($R_\mathrm{Petro}$ RMSE $\approx$ 0.4--0.8\,dex for $M_\star < 10^{11}\,\mathrm{M_\odot}$), while EAGLE underproduces them ($R_\mathrm{50}$ RMSE $\approx$ 0.4--1\,dex). TNG50 generally shows an excess of large galaxies according to both metrics; TNG100 exhibits similar but milder excess. These persistent mismatches suggest that the physical processes governing galaxy sizes---especially gas accretion histories, angular momentum redistribution, and feedback efficiency---remain inadequately captured in current cosmological simulations.

The projected ellipticity distributions (Fig.~4) show improved but still highly significant tensions. TNG performs best (RMSE $\sim0.2$--$0.4$\,dex), whereas SIMBA exhibits the largest discrepancies (RMSE $\sim0.6$--$0.9$\,dex). Yet all three simulations share a common failure: a deficit of highly flattened systems ($\epsilon > 0.6$). For TNG50, this deficit is most pronounced among quiescent galaxies, while star-forming galaxies are more abundant at all $\epsilon$. EAGLE reproduces the ellipticity distributions reasonably well for $\epsilon < 0.6$ in the $M_\star = 10^{10-10.5}\,\mathrm{M_\odot}$ bin. The systematic discrepancy in producing and/or maintaining highly flattened galaxies may reflect a combination of limited effective resolution and angular momentum loss during disk formation, as well as subsequent dynamical heating and thickening from mergers, feedback‑driven perturbations, and numerical effects \citep{Ludlow+19, Pillepich+19, vandeSande+19, Haslbauer+22, YuS+26}.

In addition, none of the simulations reproduce the observed asymmetry (Fig.~5) or concentration (Fig.~6) index distributions. The discrepancies exceed $5\sigma$ in nearly all subsamples. Notably, TNG50---despite its higher resolution---produces galaxies that are systematically more asymmetric than observed, with RMSE values $\sim0.7-1.0$\,dex. For concentration, TNG100 generally shows better agreement with observations than TNG50, which shows an excess of galaxies by a factor of $\sim2$ at the highest masses, particularly at the high-concentration end. Conversely, SIMBA exhibits a pronounced deficit of high‑concentration systems across all subsamples, with RMSE values ranging from 0.3 to 0.6\,dex depending on stellar mass. EAGLE underproduces concentrated quiescent galaxies, especially at $M_\star > 10^{10.5}\,\mathrm{M_\odot}$ (RMSE $\approx 0.7$--$0.9$\,dex).

\section{Discussion}

The consistency of these structural discrepancies across independent simulations points to shared limitations in current galaxy formation models. Increasing numerical resolution---as demonstrated by TNG50---does not fully resolve these tensions. Indeed, the ordering in Fig.~1 cannot be explained simply by resolution: TNG50 and TNG100 differ by a factor of $\sim16$ in baryonic mass resolution yet are overall much closer to the observed high-mass $n$ distributions than EAGLE and SIMBA, whereas EAGLE has a baryonic mass resolution similar to TNG100 but exhibits a deficit of high-$n$ quiescent systems more akin to SIMBA. Moreover, some of the largest offsets occur in the highest-mass bin, where particle sampling is best. We therefore regard resolution as an important differential contributor, rather than a unique explanation, with the remaining differences also reflecting the implemented galaxy-formation physics and resulting intrinsic stellar structures. While cosmic variance and finite-volume effects may contribute to the observed differences, they are unlikely to fully account for the overall discrepancies.

These tensions likely arise from a combination of subgrid treatments of star formation, feedback, and angular momentum redistribution, as well as numerical effects and differences in baryon-driven galaxy evolutionary histories. Numerical effects also contribute: the simulations employ relatively massive dark matter particles, which interact too strongly with dynamically cold stellar components, leading to spurious heating and artificially thickened disks \citep{Ludlow+19}. This effect is particularly severe in low-mass halos but accumulates across all halos over cosmic time. Additionally, the imposed gas temperature floor ($\sim 10^4$\,K) prevents the formation of dense, cold gas, producing puffier star-forming regions and limiting the growth of compact stellar bulges. In TNG, kinetic feedback further decouples gas from inflows near galaxy centers, suppressing nuclear stellar densities. Targeted tests of feedback strength and coupling could verify whether these processes drive the discrepancies. Nevertheless, the differences between TNG50 and TNG100, which share identical subgrid physics but differ in resolution, demonstrate that numerical resolution also plays an important role.

Beyond these internal modeling uncertainties, some discrepancies may also reflect uncertainties in the hierarchical assembly of galaxies, particularly through low-mass satellites and merger-driven growth. Because massive galaxies assemble a substantial fraction of their stellar mass through the accretion of lower-mass satellites, inaccuracies in the structure, star-formation histories, gas content, and quenching of low-mass systems can propagate into the predicted structural properties of massive galaxies. Existing comparisons between observed and simulated merger populations---including merger rates, mass ratios, and gas fractions---remain limited. In addition, low-mass galaxies remain challenging to model consistently across environments and cosmic time; in dense regions, low-mass satellites are often over-quenched \citep{YesufBottrell26}. These combined physical and numerical limitations lead to incorrect structure, demonstrating that SAFs provide a stringent, multidimensional test of simulations beyond average scaling relations.

It could be argued that the discrepancies identified here do not constitute a fundamental challenge to galaxy formation theory because current models involve a large number of coupled parameters, substantial uncertainties, and complex physical prescriptions that remain imperfectly constrained. From this perspective, the observed tensions may simply reflect the fact that existing simulations are still too crude to reproduce structural abundance functions at the level of accuracy now achievable with current observations, and that future improvements in galaxy formation models may reduce these discrepancies. However, this interpretation does not weaken our central conclusion: different state-of-the-art simulations disagree with observations in different ways, indicating that they do not yet provide a sufficiently robust framework for interpreting galaxy evolution to the extent often assumed. The longstanding strategy of calibrating simulations using a limited set of observables, such as the stellar mass function, has not yet produced a framework capable of simultaneously reproducing a broader range of galaxy properties. Both this work and our recent study \citep{YesufBottrell26} reveal substantial discrepancies across multiple observables, highlighting a key challenge for galaxy formation simulations: developing a self-consistent and predictive framework that can simultaneously reproduce the observed structural and subpopulation abundances, black hole demographics, and environmental trends in star formation, while providing deeper physical insight into the processes governing galaxy evolution.

Looking ahead, the systematic structural discrepancies revealed by our univariate analysis across multiple parameters at fixed stellar mass motivate a natural extension of the SAF framework to joint multivariate distributions of galaxy structure, stellar mass, and star-formation rate. Such an approach would enable identification of specific regions of parameter space where simulations fail---for example, systems with realistic galaxy sizes but incorrect morphologies---thereby providing stronger constraints on the physical processes governing structural evolution and quenching. Extending these comparisons to lower stellar masses ($\log(M_\star/M_\odot) < 10$) and higher redshifts ($z \sim 2$) will further determine whether the structural discrepancy evolves with cosmic time and help disentangle the relative roles of stellar and AGN feedback. Although constructing these joint distributions is beyond the scope of the present work, they represent a critical next step toward establishing robust benchmarks for testing galaxy formation models against the full diversity of observed galaxy structure. Ultimately, multivariate SAFs provide a compact and physically interpretable description of large galaxy survey data, analogous to the role of the stellar mass function in studies of galaxy evolution, while establishing a new standard for validating next-generation cosmological simulations. 

\section{Conclusions}

We present a uniform, population‑level census of galaxy structural abundance functions (SAFs), providing a stringent test of the current generation of cosmological hydrodynamical simulations. By moving beyond average scaling relations to evaluate the full distribution of galaxy structures, we uncover systematic and statistically significant discrepancies across all examined parameters: S\'{e}rsic index, half‑light and Petrosian radii, ellipticity, asymmetry, and concentration. Critically, none of the flagship simulations---EAGLE, IllustrisTNG, or SIMBA---simultaneously reproduce the observed structural distributions across the full range of stellar mass and star‑formation activity.

These discrepancies expose a significant challenge for current galaxy formation simulations. While individual simulations display diverse shortcomings---underproducing or overproducing compact spheroids and extended or round galaxies---all underproduce highly flattened disks. The persistence of these offsets even in the higher‑resolution TNG50 run demonstrates that the problem is not primarily driven by numerical resolution alone, but instead may reflect shortcomings in the modelling of the baryon cycle---including star formation, feedback, and angular momentum redistribution---as well as associated numerical effects.

Our results establish that agreement with the global stellar mass function is necessary but not sufficient for realistic galaxy formation. While simulations are calibrated to reproduce integrated stellar mass statistics, they differ markedly in how that mass is spatially distributed, as revealed by SAFs. Even when a model performs well in some regimes, it fails in others, systematically misrepresenting key structural metrics. This provides clear evidence that models can be ``right for the wrong reasons'', with similar bulk outcomes arising from divergent physical pathways that are exposed only through multidimensional structural analyses and subpopulation‑level comparisons.

As we enter the era of complementary, high‑quality surveys utilising JWST (deep, high‑resolution, small‑area), the Chinese Space Station Telescope (CSST) and Euclid (wide‑field space‑based mapping), and the Vera Rubin Observatory (wide‑area, ground‑based imaging), the SAF framework offers a sensitive, population‑level benchmark for the next generation of galaxy formation models, applicable across cosmic time. Incorporating these structural constraints into model testing---and, where appropriate, calibration---will be essential for achieving not only the correct stellar mass distributions but also realistic galaxy structures. While some level of simplification is unavoidable in numerical models, integrating SAFs into evaluation frameworks enables more robust inference on how galaxies assemble their mass and on the physical processes that govern its distribution within and around galaxies.

\begin{figure*}
\plotone{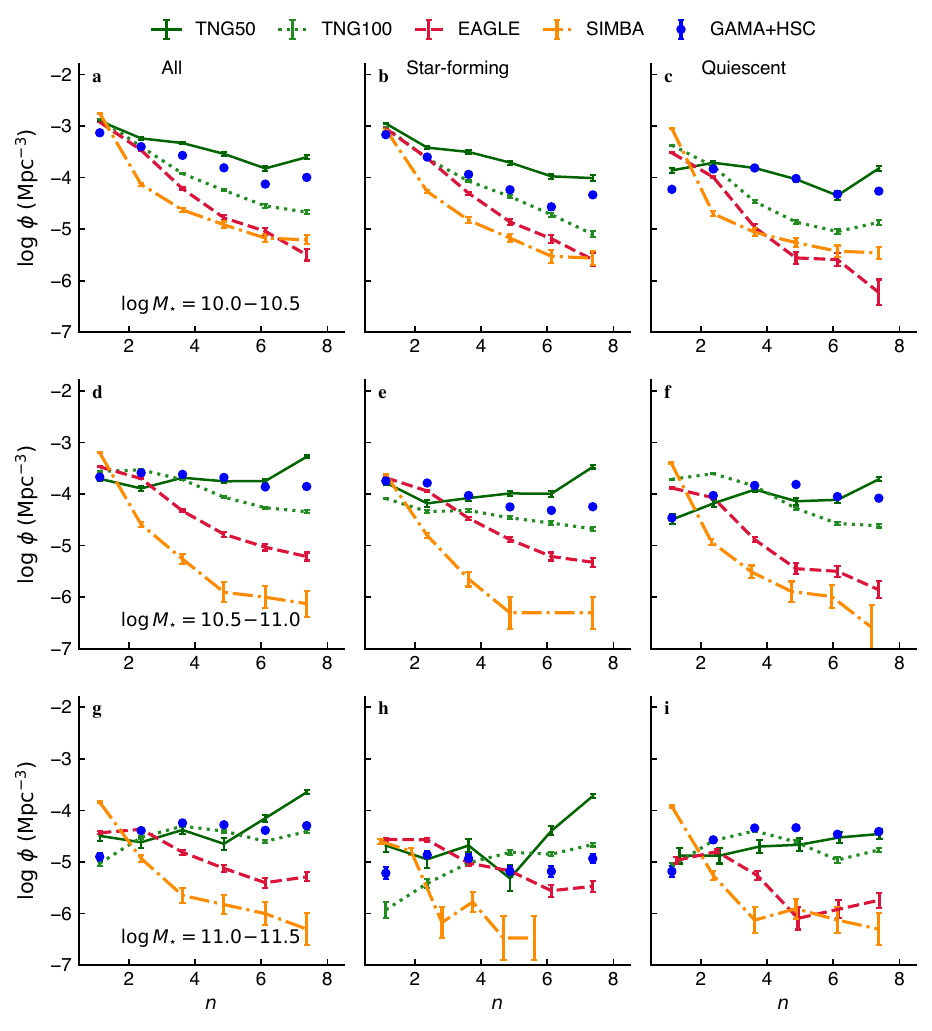}
\caption{\textbf{S\'{e}rsic index function.}  
Number densities of galaxies as a function of i-band Sersic index, $n$, shown in panels for three stellar mass bins ($\log M_\star / M_\odot = 10.0$--$10.5$, $10.5$--$11.0$, $11.0$--$11.5$) and three star formation categories (All galaxies, Star-forming: $\Delta \log \mathrm{SFR} > -1$, Quiescent: $\Delta \log \mathrm{SFR} < -1$). Each panel displays the logarithmic number density, $\log \phi(n)\;[\mathrm{Mpc^{-3}}]$, versus S\'{e}rsic index.  Simulations are indicated by colored lines: TNG50 (dark green, solid), TNG100 (forest green, dotted), EAGLE (crimson, dashed), and SIMBA (dark orange, dash-dotted). Observational data from HSC+GAMA are shown as blue circles.}
\label{fig:SernF}
\end{figure*}

\begin{figure*}
\plotone{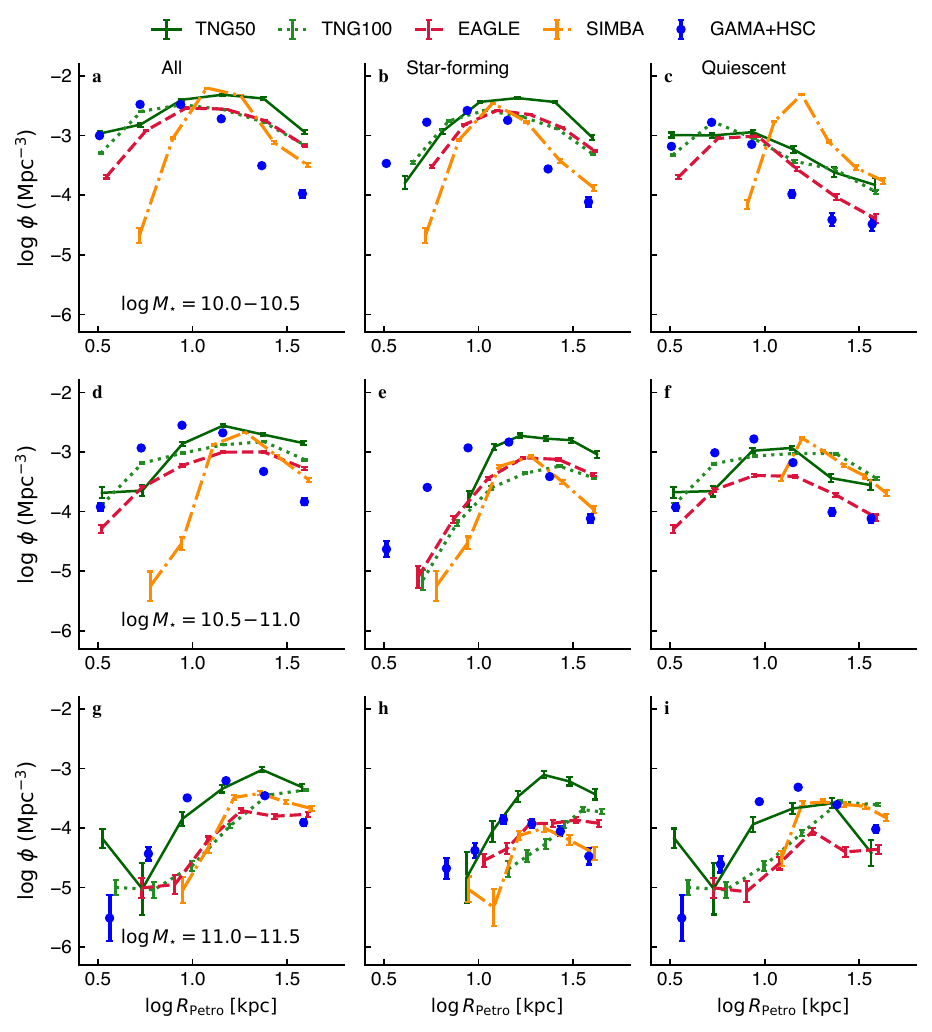}
\caption{\textbf{Galaxy size function.}  
Number densities of galaxies as a function of logarithmic Petrosian radius, $\log\,R_\mathrm{Petro}\,[\mathrm{kpc}]$, shown in panels for three stellar mass bins ($\log M_\star / M_\odot = 10.0$--$10.5$, $10.5$--$11.0$, $11.0$--$11.5$) and three star formation categories 
(All galaxies, Star-forming: $\Delta \log \mathrm{SFR} > -1$, Quiescent: $\Delta \log \mathrm{SFR} < -1$). Each panel displays the logarithmic number density, $\log \phi(R_\mathrm{Petro})\;[\mathrm{Mpc^{-3}}]$, versus galaxy size.  Simulations are indicated by colored lines: TNG50 (dark green, solid), 
TNG100 (forest green, dotted), EAGLE (crimson, dashed), and 
SIMBA (dark orange, dash-dotted). Observational data from HSC+GAMA are shown as blue circles.}
\label{fig:RpF}
\end{figure*}

\begin{figure*}
\plotone{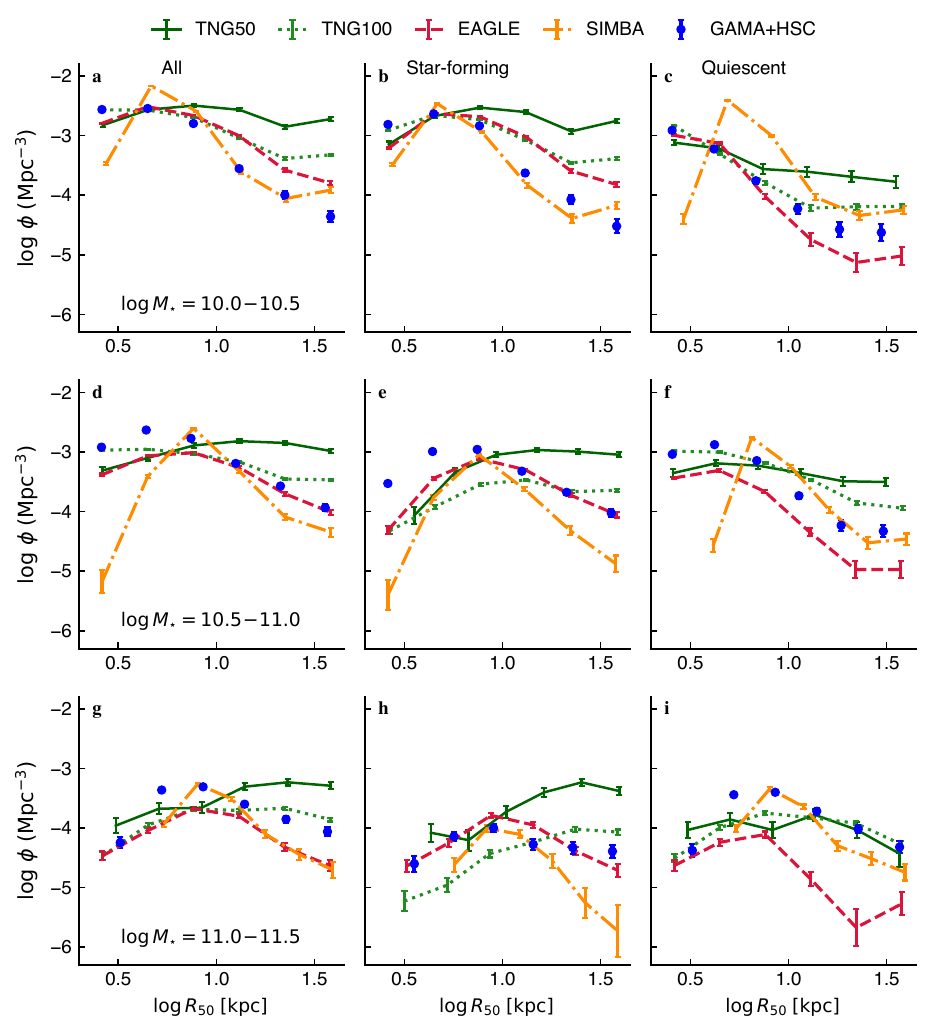}
\caption{\textbf{Galaxy half-light size function.}  
Number densities of galaxies as a function of logarithm of i-band S\'{e}rsic half-light radius, $\log\,R_\mathrm{50}\,[\mathrm{kpc}]$, shown in panels for three stellar mass bins ($\log M_\star / M_\odot = 10.0$--$10.5$, $10.5$--$11.0$, $11.0$--$11.5$) and three star formation categories (All galaxies, Star-forming: $\Delta \log \mathrm{SFR} > -1$, Quiescent: $\Delta \log \mathrm{SFR} < -1$). Each panel displays the logarithmic number density, $\log \phi(R_\mathrm{50})\;[\mathrm{Mpc^{-3}}]$, versus galaxy size.  Simulations are indicated by colored lines: TNG50 (dark green, solid), 
TNG100 (forest green, dotted), EAGLE (crimson, dashed), and 
SIMBA (dark orange, dash-dotted). Observational data from HSC+GAMA are shown as blue circles.}
\label{fig:R50F}
\end{figure*}

\begin{figure*}
\plotone{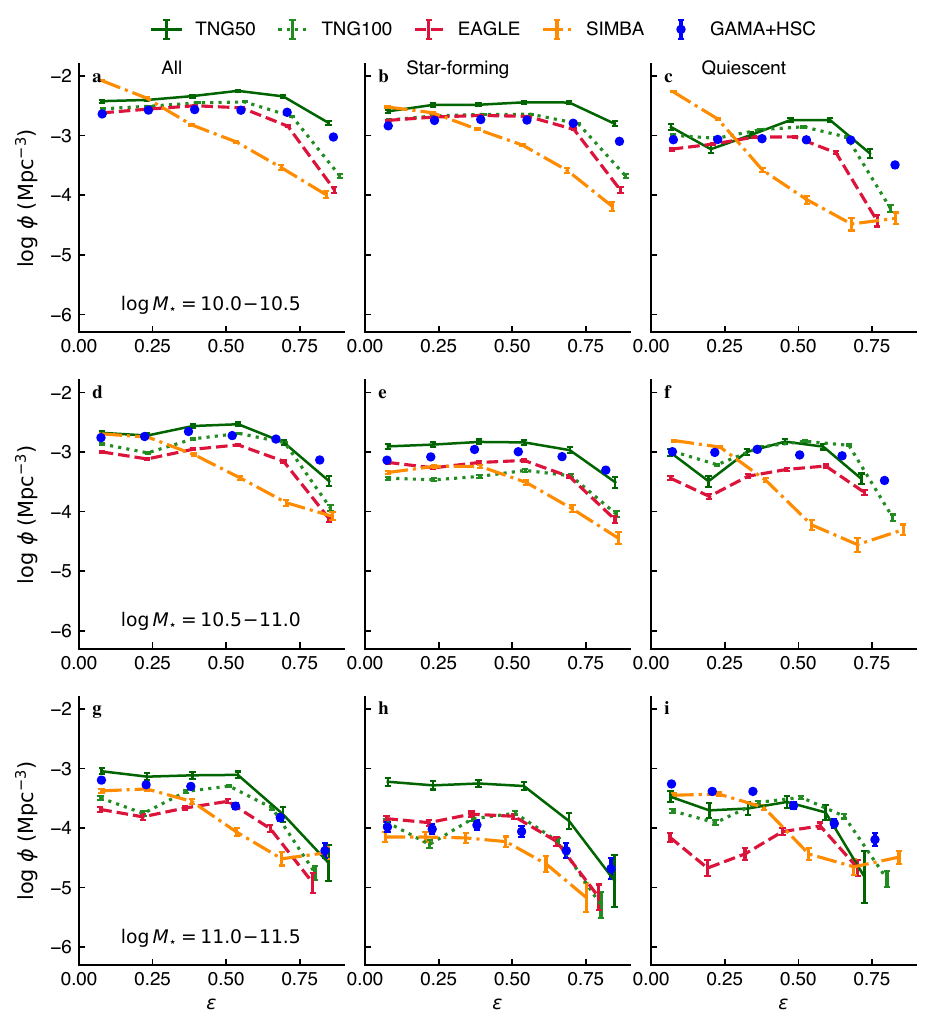}
\caption{\textbf{Galaxy ellipticity function.} Number densities of galaxies as a function of ellipticity $\epsilon$, shown in panels for three stellar mass bins ($\log M_\star / M_\odot = 10.0$--$10.5$, $10.5$--$11.0$, $11.0$--$11.5$) and three star formation categories (All galaxies, Star-forming: $\Delta \log \mathrm{SFR} > -1$, Quiescent: $\Delta \log \mathrm{SFR} < -1$). Each panel shows the logarithmic number density, $\log \phi(\epsilon)\;[\mathrm{Mpc^{-3}}]$, versus ellipticity. Simulations are indicated by colored lines: TNG50 (dark green, solid), TNG100 (forest green, dotted), EAGLE (crimson, dashed), and SIMBA (dark orange, dash-dotted). Observational data from HSC+GAMA are shown as blue circles.}
\label{fig:EllipF}
\end{figure*}

\begin{figure*}
\plotone{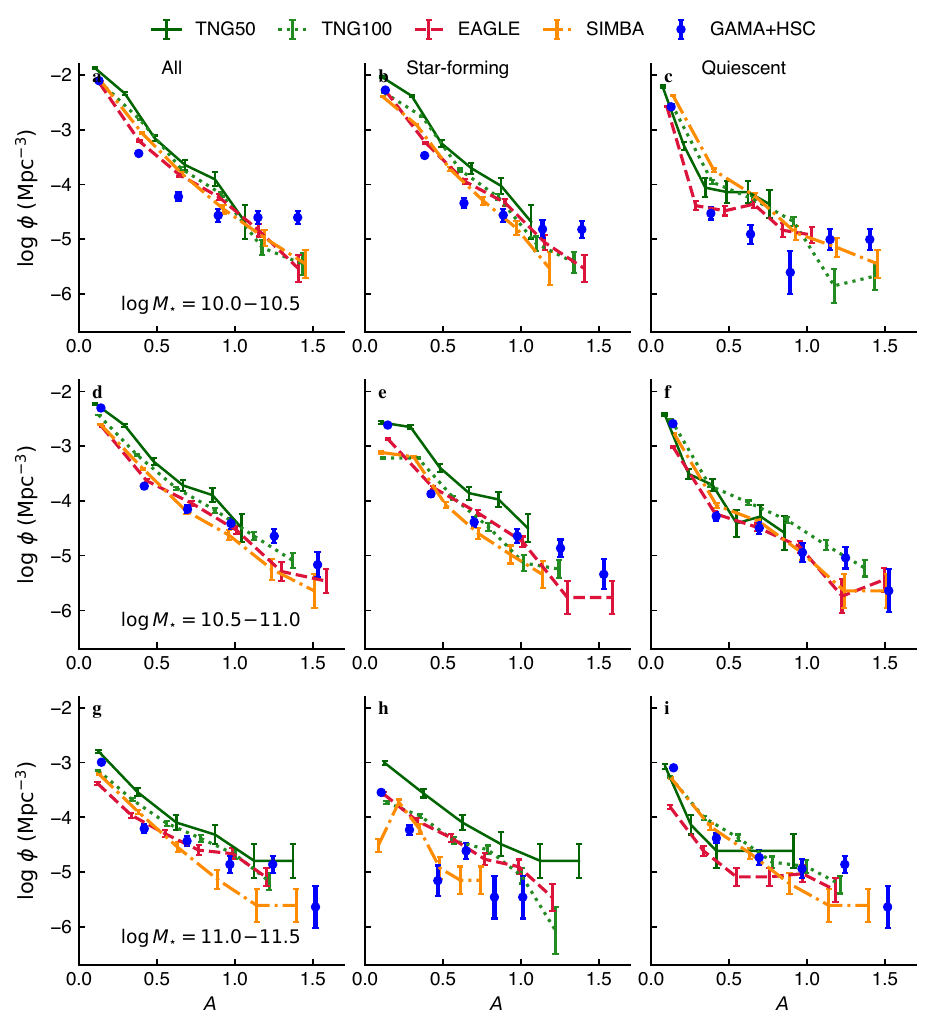}
\caption{\textbf{Structural Asymmetry function.}  
Number densities of galaxies as a function of i-band asymmetry, $A$, shown in panels for three stellar mass bins ($\log M_\star / M_\odot = 10.0$--$10.5$, $10.5$--$11.0$, $11.0$--$11.5$) and three star formation categories (All galaxies, Star-forming: $\Delta \log \mathrm{SFR} > -1$, Quiescent: $\Delta \log \mathrm{SFR} < -1$). Each panel displays the logarithmic number density, $\log \phi(A)\;[\mathrm{Mpc^{-3}}]$, versus asymmetry.  Simulations are indicated by colored lines: TNG50 (dark green, solid), 
TNG100 (forest green, dotted), EAGLE (crimson, dashed), and 
SIMBA (dark orange, dash-dotted). Observational data from HSC+GAMA are shown as blue circles.}
\label{fig:AsymF}
\end{figure*}

\begin{figure*}
\plotone{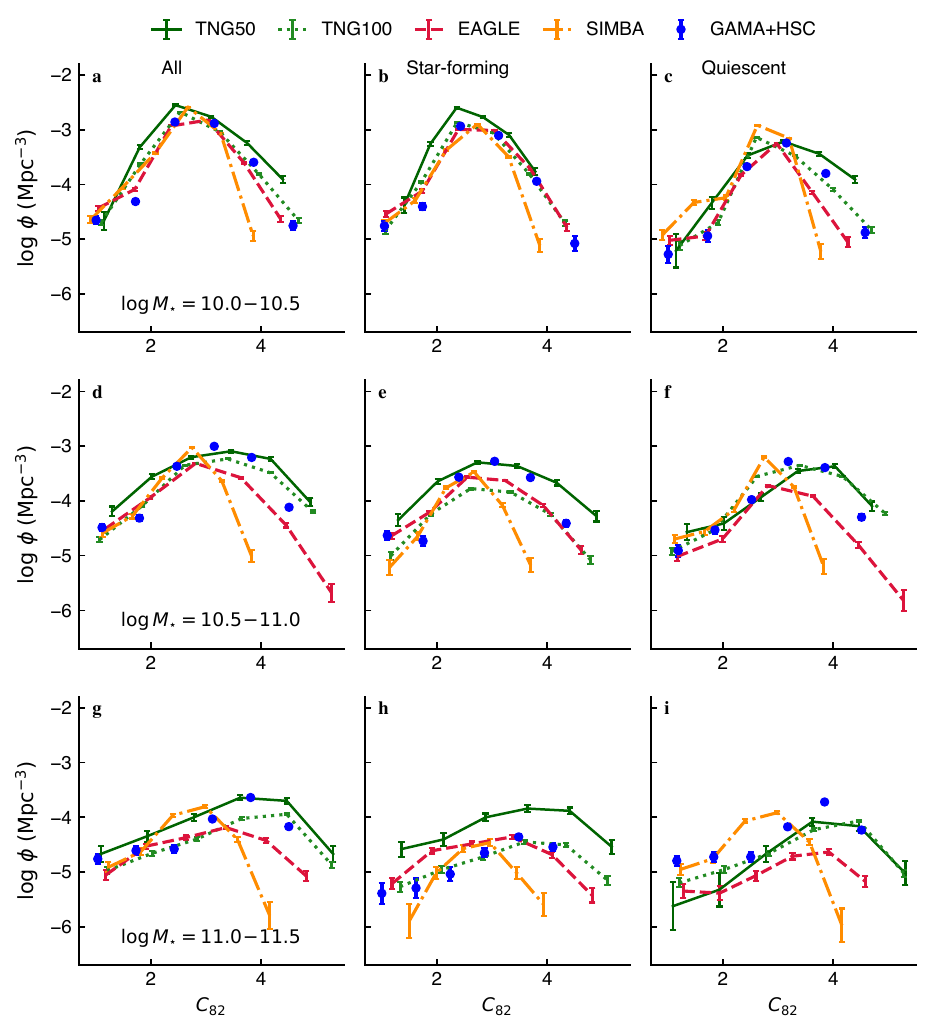}
\caption{\textbf{Galaxy concentration function.} Number densities of galaxies as a function of concentration $C_{82}$, shown in panels for three stellar mass bins ($\log M_\star / M_\odot = 10.0$--$10.5$, $10.5$--$11.0$, $11.0$--$11.5$) and three star formation categories (All galaxies, Star-forming: $\Delta \log \mathrm{SFR} > -1$, Quiescent: $\Delta \log \mathrm{SFR} < -1$). Each panel shows the logarithmic number density, $\log \phi(C)\;[\mathrm{Mpc^{-3}}]$, versus concentration. Simulations are indicated by colored lines: TNG50 (dark green, solid), TNG100 (forest green, dotted), EAGLE (crimson, dashed), and SIMBA (dark orange, dash-dotted). Observational data from HSC+GAMA are shown as blue circles.}
\label{fig:C82F}
\end{figure*}

\medskip

We sincerely thank the anonymous referees for their useful and constructive feedback, which has significantly improved the manuscript. We thank M. Reza Ayromlou for providing simulation data. C.B. gratefully acknowledges support from the Forrest Research Foundation. H.M.Y. was partially supported JSPS KAKENHI Grant Number JP22K14072. This work was supported by resources provided by the Pawsey Supercomputing Research Centre’s Setonix Supercomputer (https://doi.org/10.48569/18sb-8s43) and Acacia Object Storage (https://doi.org/10.48569/nfe9-a426), with funding from the Australian Government and the Government of Western Australia.

\medskip

H.M.Y. conceived the project, carried out the data analysis, and wrote the initial manuscript draft. C.B. performed the structural measurements, which were reviewed and validated by H.M.Y. H.M.Y., C.B., and L.C.H. developed the main interpretation. A.S.G.R. contributed an R template for fitting with ProFuse and advised on the ProFuse output data model. S.B. contributed to calibration and fine-tuning of ProFound settings for HSC images. R.H.W.C. contributed to development of our ProFuse pipeline and the image stacking procedure for ProFound. All authors contributed to the revision and editing of the manuscript.

\clearpage

\appendix
\renewcommand{\thefigure}{A\arabic{figure}}
\setcounter{figure}{0}

\section{Appendix information}

\begin{figure}
\plotone{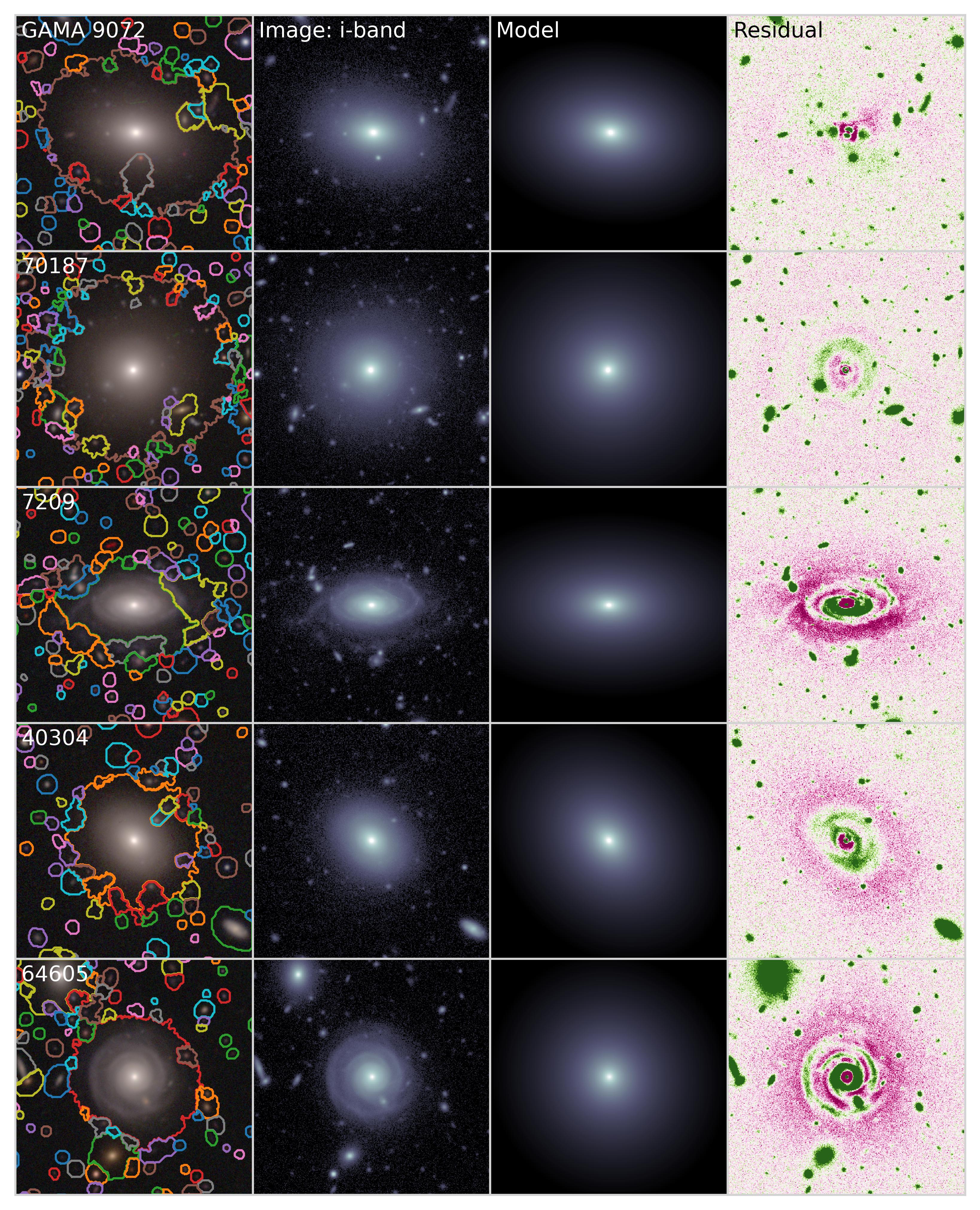}
\caption{\textbf{Examples of bright HSC galaxies and their structural modelling.} \textbf{a}, HSC $gri$ colour composites, including masks for foreground and background sources. \textbf{b}, HSC $i$-band images used for the structural analysis. \textbf{c}, Best-fitting $i$-band ProFit models. \textbf{d}, Residual images obtained by subtracting the models from the data. Residual colour scales span $-5$ to $+5$ times the average sky noise, where purple indicates over-subtracted regions and green indicates under-subtracted regions. For visualization purposes, the displayed images are zoomed by a factor of three relative to those used for source detection and deblending, enhancing the visibility of the galaxy light distribution and residual structures. \label{fig:HSCcutoutBright}}
\end{figure}

\begin{figure}
\plotone{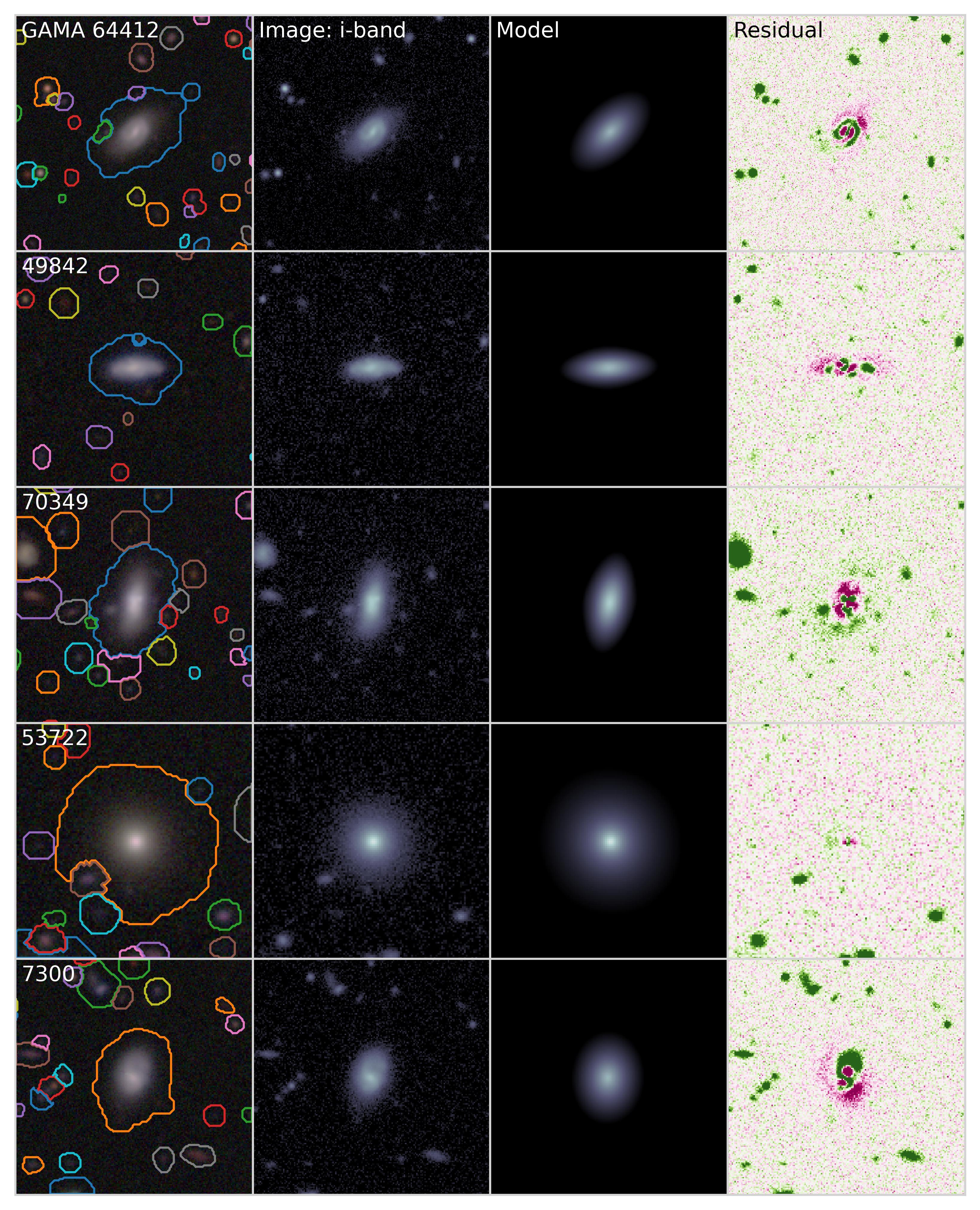}
\caption{\textbf{Examples of faint HSC galaxies and their structural modelling.} \textbf{a}, HSC $gri$ colour composites, including masks for foreground and background sources. \textbf{b}, HSC $i$-band images used for the structural analysis. \textbf{c}, Best-fitting $i$-band structural models. \textbf{d}, Residual images obtained by subtracting the models from the data. Residual colour scales span $-5$ to $+5$ times the average sky noise, where purple indicates over-subtracted regions and green indicates under-subtracted regions. The images are displayed at three times the scale used for source detection and deblending for clarity. \label{fig:HSCcutoutFaint}}
\end{figure}

\begin{figure}
\plotone{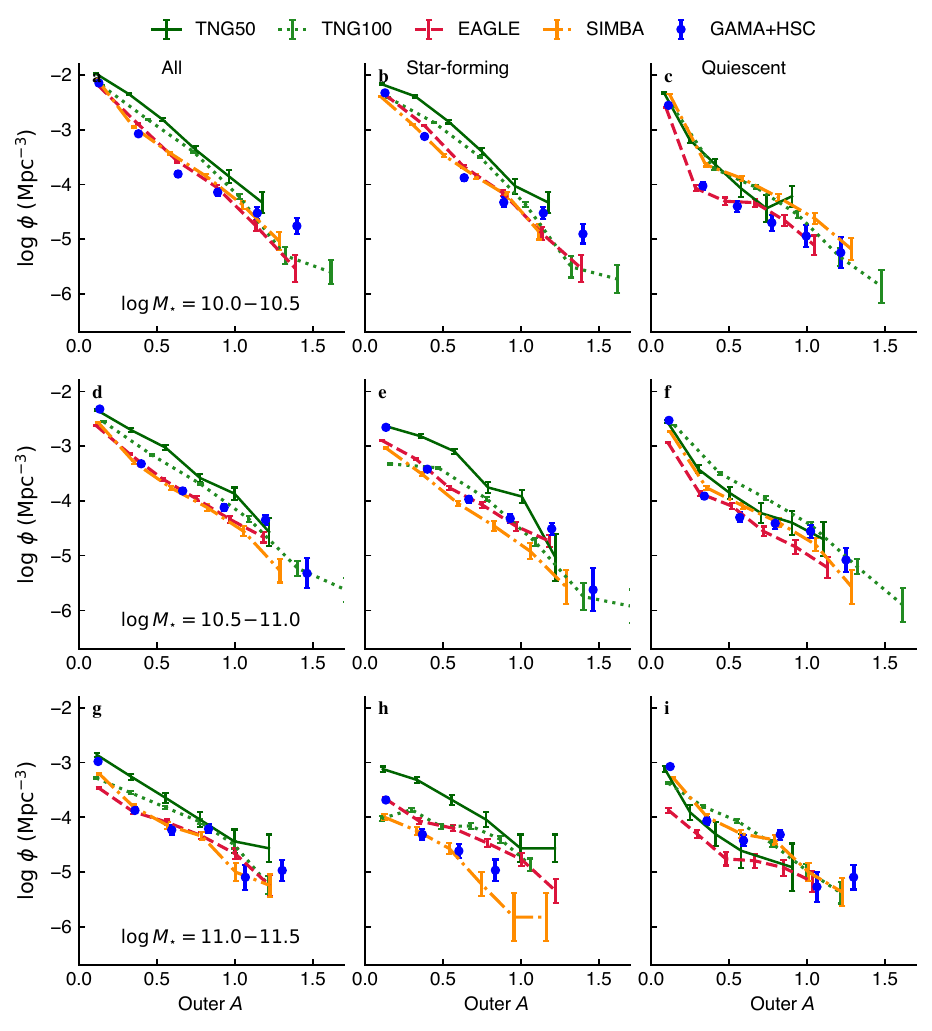}
\caption{\textbf{Outer Asymmetry function.} Number densities of galaxies as a function of i-band outer asymmetry, shown in panels for three stellar mass bins ($\log M_\star / M_\odot = 10.0$--$10.5$, $10.5$--$11.0$, $11.0$--$11.5$) and three star formation categories (All galaxies, Star-forming: $\Delta \log \mathrm{SFR} > -1$, Quiescent: $\Delta \log \mathrm{SFR} < -1$).}
\label{fig:OutAsymF}
\end{figure}

\begin{figure}
\plotone{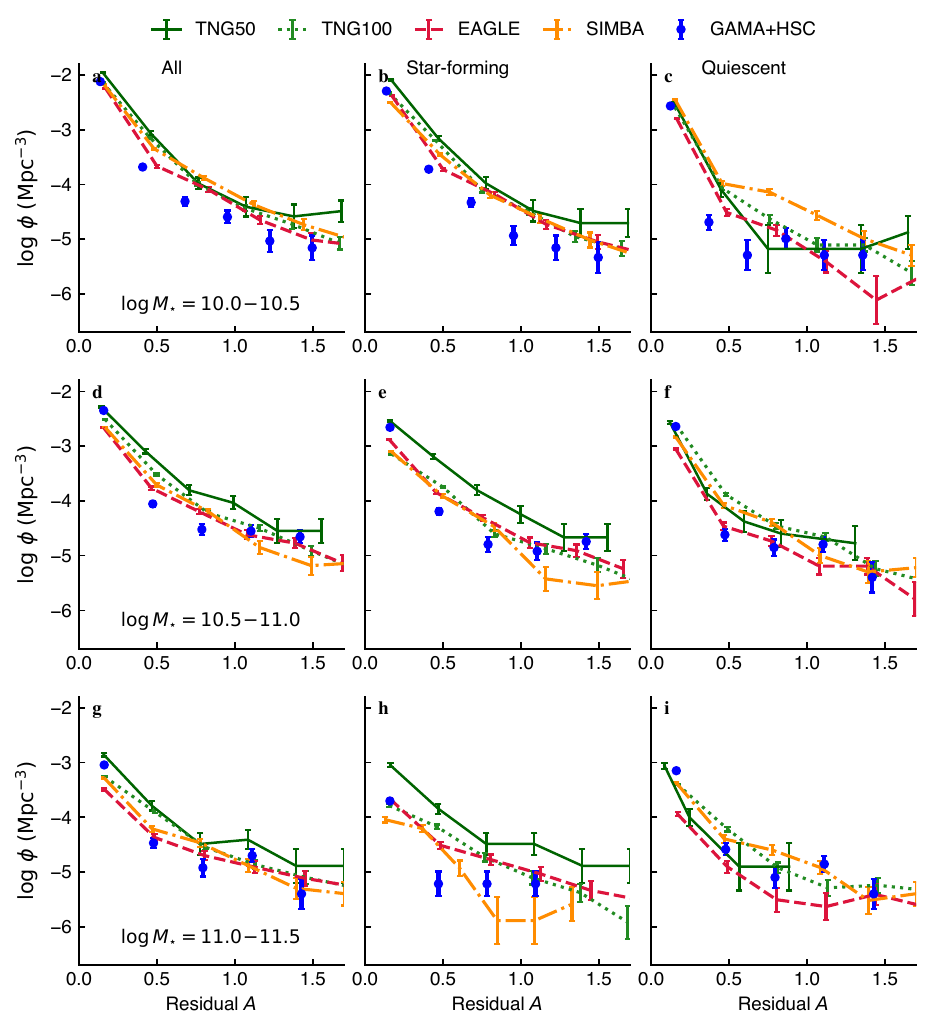}
\caption{\textbf{Residual Asymmetry function.} The same as the main figure 6, except $A$ is measured after subtracting smooth S\'{e}rsic model.}
\label{fig:RMSAsymF}
\end{figure}

\begin{figure}
\plotone{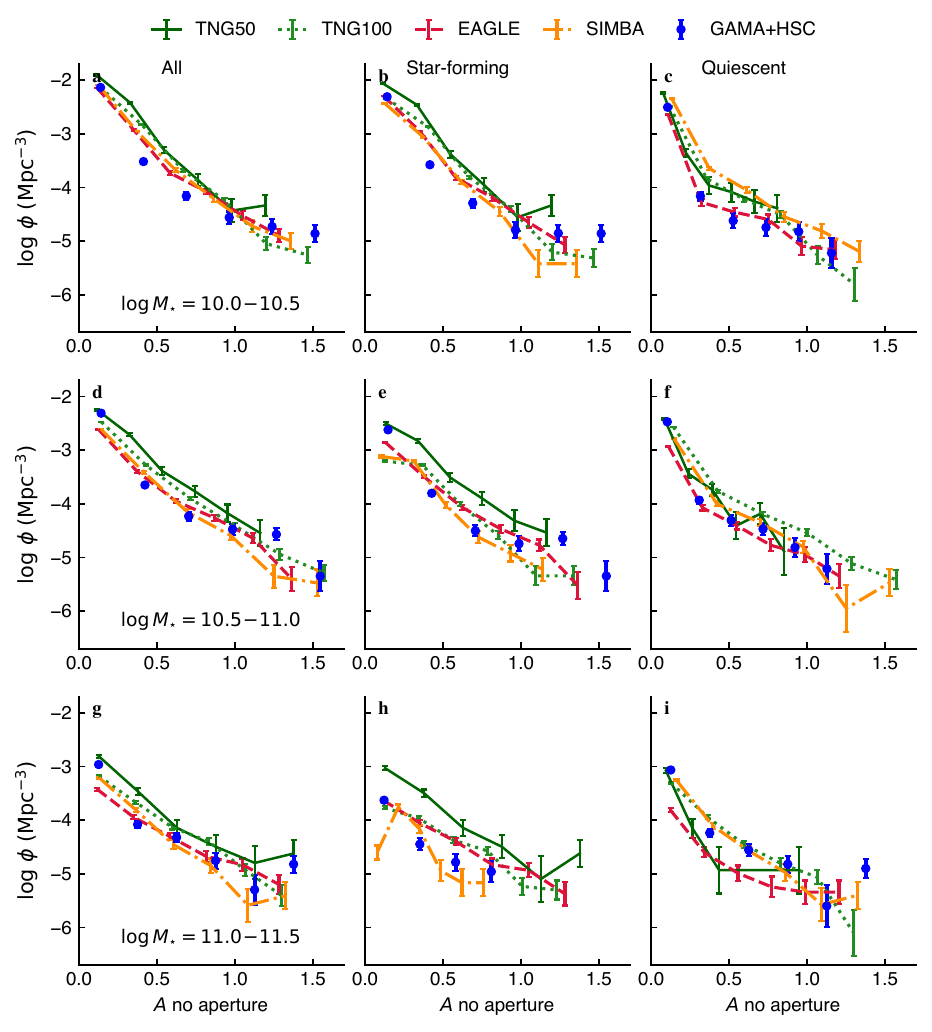}
\caption{\textbf{No Aperture Asymmetry function.} The same as the main figure 6, except $A$ is measured in the entire segmentation map without restricting the aperture to $r = 1.5R_p$.}
\label{fig:NoApAsymF}
\end{figure}

\begin{figure}
\plotone{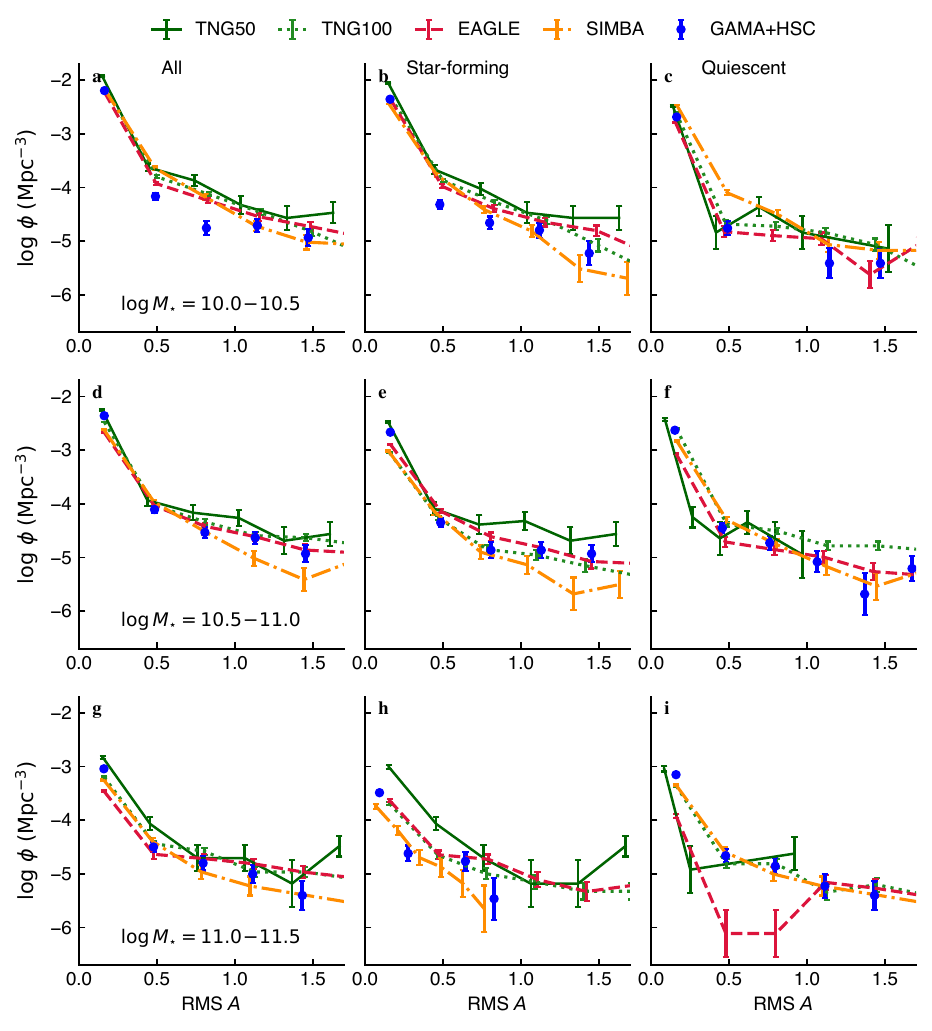} 
\caption{\textbf{RMS Asymmetry function.} Root-mean-squared asymmetry is an alternative definition $A$, where the residual flux between image and its rotated counterpart is squared rather than absolute-valued. }
\label{fig:NoApAsymF}
\end{figure}


\bibliographystyle{aasjournalv7}
\end{document}